\documentclass[12pt]{article}
\usepackage[utf8]{inputenc}
\usepackage{graphicx}
\usepackage{amsmath}
\usepackage{hyperref}
\usepackage{geometry}
\usepackage{cite}
\usepackage{multirow}
\usepackage{float}

\title{Breast Tumor Classification Using EfficientNet Deep Learning Model}
\author{
    Majid Behzadpour\thanks{Department of Electrical and Computer Engineering, University of Tehran, Tehran, Iran. Email: majid.behzadpour11@gmail.com} \and 
    Bengie L. Ortiz\thanks{Department of Pediatrics, Hematology and Oncology Division, Michigan Medic
    ine, University of Michigan Health System, Ann Arbor, MI, USA.} \and 
    Ebrahim Azizi\thanks{Department of Electrical and Computer Engineering, Texas Tech University, Lubbock, TX, USA.} \and 
    Kai Wu\thanks{Department of Electrical and Computer Engineering, Texas Tech University, Lubbock, TX, USA. Email: kai.wu@ttu.edu}
}

\date{}

\begin{document}

\maketitle

\begin{abstract}
Precise breast cancer classification on histopathological images has the potential to greatly improve the diagnosis and patient outcome in oncology. The data imbalance problem largely stems from the inherent imbalance within medical image datasets, where certain tumor subtypes may appear much less frequently. This constitutes a considerable limitation in biased model predictions that can overlook critical but rare classes. In this work, we adopted EfficientNet, a state-of-the-art convolutional neural network (CNN) model that balances high accuracy with computational cost efficiency. To address data imbalance, we introduce an intensive data augmentation pipeline and cost-sensitive learning, improving representation and ensuring that the model does not overly favor majority classes. This approach provides the ability to learn effectively from rare tumor types, improving its robustness. Additionally, we fine-tuned the model using transfer learning, where weights in the beginning trained on a binary classification task were adopted to multi-class classification, improving the capability to detect complex patterns within the BreakHis dataset. Our results underscore significant improvements in the binary classification performance, achieving an exceptional recall increase for benign cases from 0.92 to 0.95, alongside an accuracy enhancement from 97.35 \% to 98.23\%. Our approach improved the performance of multi-class tasks from 91.27\% with regular augmentation to 94.54\% with intensive augmentation, reaching 95.04\% with transfer learning. This framework demonstrated substantial gains in precision in the minority classes, such as Mucinous carcinoma and Papillary carcinoma, while maintaining high recall consistently across these critical subtypes, as further confirmed by confusion matrix analysis.
\end{abstract}


\section{Introduction}
Breast cancer is one of the most lethal diseases, and its prevalence is rapidly increasing in both developed and developing nations \cite{who_mammography}. Among various forms of cancers, breast cancer has the lowest survival rate \cite{miller2022cancer}, which makes this health problem highly critical. Accurate and early diagnosis is important for efficient treatment. Medical images play a crucial role in modern healthcare by providing valuable information that aids in early detection, thereby increasing the survival rate. However, manual interpretation of these images is time-consuming, and the accuracy of detection depends on the pathologist’s experience. This process is further complicated by the complexity of interpreting images from a wide variety of medical imaging technologies \cite{zheng2007enhancing}. In line with this, various automation techniques for disease diagnosis on different types of medical images are currently under development. Various machine learning-based approaches, such as Naive Bayes and Support Vector Machines, have been proposed to develop efficient systems with high accuracy for diagnosing abnormalities \cite{lopez2020clinical}, \cite{aswathy2021svm}. However, traditional machine learning models struggle to handle large, complex datasets due to their reliance on manual feature extraction. Researchers, therefore, favor deep learning models due to their capability of automatic feature extraction \cite{rana2023machine}.  Various deep learning-based models are being proposed due to their success in breast cancer classification (BCC). Convolutional Neural Networks (CNNs) have significantly improved classification accuracy, reaching rates as high as 95.64\% \cite{al2018fully}. This suggests that CNNs are more reliable diagnostic tools than traditional machine learning methods, as they can automatically extract subtle features and patterns from large datasets, enhancing the system’s robustness and accuracy. The implementation of AlexNet \cite{krizhevsky2017imagenet} with fine-tuning strategies \cite{nawaz2018classification} marked a significant advancement, followed by the Visual Geometry Group (VGG) network, which, combined with handcrafted feature extraction methods \cite{al2023analyzing}, \cite{simonyan2014very}, demonstrated substantial improvements in BCC performance. The Inception architecture \cite{alkassar2021going} further advanced the field by combining convolutional filters of varying sizes within a single layer. Numerous modified models based on these architectures have been introduced for BCC, as discussed in Section 2.
In this work, we have employed the EfficientNet model \cite{tan2019efficientnet}, representing a state-of-the-art architecture in CNN. EfficientNet is characterized by its high accuracy and low computational complexity. It systematically scales up key architectural dimensions by adjusting the depth (number of layers), width (number of filters per layer), and resolution of the input images. This approach achieves a good balance between performance and computational cost, so it is quite appropriate for applications like medical image analysis, making the models feasible for use in real-time applications and resource-limited healthcare settings.
One of the critical issues in neural network training is data imbalance, particularly in domains like medical image classification, where certain classes (medical images) representing specific tumors are underrepresented compared to others. These imbalanced classes cause neural networks to be biased toward the majority classes, leading to poor generalization. This imbalance affects not only the accuracy of these models but also diminishes their ability to detect rare or less frequent classes. This presents a significant challenge in medical applications where each class, regardless of the rarity, can carry critical clinical importance \cite{chen2022personalized}. For example, in the classification of histopathological images, an imbalanced class distribution can result in the neglect of some very critical and rare cancer subtypes, hence undermining the effectiveness of a diagnostic tool. Addressing this imbalance is therefore essential to ensure unbiased predictions by the neural network across all classes, making it a critical factor in developing robust, clinically useful models. This work develops a robust model to classify histopathology images from the BreaKHis dataset \cite{spanhol2015dataset}. Our primary goals are to improve classification accuracy, address class imbalance, and enhance the interpretability of the model’s predictions.
This paper is arranged as follows: In Section 2, we review related works that provide a foundation for our approach. Section 3 details our methodology, including the dataset, model architecture, optimization techniques, and training process. In Section 4, we present the classification results and discuss their implications, followed by a comparison of our findings with existing works to highlight the efficiency of our proposed EfficientNet deep learning model.

\section{Related Works}
Most of the current research on deep learning for BCC focuses on using advanced neural network architectures like Xception, DenseNet, and VGGNet, typically combined with data augmentation techniques to handle imbalanced datasets and improve classification accuracy. Several works have underlined the use of transfer learning and generative models, such as generative adversarial network (GAN), in enhancing representational features of underrepresenting classes, obtaining high precision and recall rates. Additionally, hybrid models that combine CNNs with techniques like Long Short Term Memory (LSTM) and Extreme Gradient Boosting known as XGBoost demonstrate average accuracies of 91.60\% and 92.50\% on different magnification levels of histopathological images \cite{maleki2023breast}, \cite{srikantamurthy2023classification}. While the binary classification often reaches higher accuracy, the multi-class classification remains challenging. Recently, competitive results have been reported utilizing advanced feature extraction and preprocessing techniques \cite{murtaza2020deep}. Handling imbalances, feature extraction, and transfer learning are considered key to achieving state-of-the-art performance, especially in multi-class classifications. Transfer learning allows leveraging pre-trained models to improve performance on small datasets, which is particularly beneficial in domains where data collection is expensive, such as medical image classification.
Abunasser et al. reported a deep learning approach toward detecting and classifying breast cancer \cite{abunasser2022breast}. They fine-tuned the Xception model for BreakHis dataset with eight classes of breast cancer and augmented by GAN. The fine-tuned Xception model was then trained, validated, and tested to achieve a precision of 97.60\%, recall of 97.60\%, and F1-score of 97.58\%. Notably, it achieved an accuracy of 96.48\% on the underrepresented malignant classes. Nawaz et al. demonstrated the performance of the DenseNet model breast cancer detection to 95.40\% validation accuracy in multi-class BCC, diagnostic performance beyond human experts \cite{nawaz2018multi}.
Houssein et al. contributed an extensive review of the latest publications in the area of breast cancer detection and classification by using different machine learning and deep learning techniques on various image modalities \cite{houssein2021deep}. Murtaza et al. contributed a comprehensive review of research in BCC, based on 49 studies \cite{murtaza2020deep}. The authors emphasized the role of public datasets in this area, particularly mammogram and histopathology images, and highlighted effective pre-processing techniques that include image augmentation and normalization. They stated that CNN models perform better compared to pre-trained models for BCC.
Reshma et al. presented a robust Genetic Algorithm involving a CNN in the process of BCC, addressing challenges in sample selection by aggregating relief-driven optimal textural, graph, and morphological features \cite{reshma2022detection}. The approach can improve the accuracy of the classification by handling low-density areas found in the feature space and demonstrating an accuracy of 92.44\% in multi-class. Bejnordi et al. proposed a pioneering deep learning system for classifying whole slide images of breast tissue biopsies \cite{bejnordi2017deep}. Their approach focuses on stromal properties as diagnostic biomarkers, demonstrating that stromal tissue alone can effectively distinguish between breast cancer and benign breast disease, their results show 92\% accuracy of classification.
Ijaz et al. provided a model of Convolutional Block Attention Module (CBAM)-VGGNet for the classification tasks related to H- and E-stained images in breast histopathology \cite{ijaz2023modality}. They trained VGG16 and VGG19 using CBAM with a Global Average Pooling (GAP) layer on datasets of cancerous histopathology, further enhancing the feature extraction pipeline. Their overall accuracy reached 98.76\%, with a 97.95\% F1-score on 400X data of the BreakHis dataset. Jaganathan et al. demonstrated the effectiveness of a transfer learning-based concatenated model by integrating pre-trained models such as VGG-16, MobileNetV2, ResNet50, and DenseNet121, achieving a 97\% validation accuracy for binary classification \cite{jaganathan2024revolutionizing}. Hameed et al. proposed a deep learning model based on six intermediate layers of the pre-trained Xception network for classifying hematoxylin-eosin-stained images of breast cancer into four categories \cite{hameed2022multiclass}. Their model fetched an accuracy of 98\% when applied to the original dataset, and 97.79\% when applying Macenko stain normalization to the dataset.
Maleki et al. proposed a method using pre-trained models to extract features, which are then classified with XGBoost \cite{maleki2023breast}. Evaluating 18 architectures on the BreakHis dataset, they achieved 93.6\%, 91.3\%, 93.8\%, and 89.0\% accuracy for 40×, 100×, 200×, and 400× magnifications, respectively. Srikantamurthy et al. developed a hybrid CNN-LSTM method using data augmentation and extracted deep convolutional features with ResNet50, InceptionV3, and CNN pre-trained on ImageNet \cite{srikantamurthy2023classification}. These features were then classified using an LSTM model, with 'adam' found to be the best optimizer. The hybrid model achieved 99\% accuracy for binary classification of benign and malignant cancer, and 92.5\% accuracy for multi-class classification of cancer subtypes. Amin et al. introduced the FabNet model, showcasing a novel approach to classifying multi-scale histopathological images by integrating hierarchical feature maps to enhance accuracy. This model achieved an average accuracy of 96.30\% in binary classification and 95.62\% in multi-class classification on the BreakHis dataset across various magnification levels \cite{amin2023fabnet}. 
Al-Jabbar et al. developed a method using Artificial Neural Network (ANN) with features selected from VGG-19 and ResNet-18 models \cite{al2023analyzing}. For multi-class datasets at 400× and 40× magnification factors, their approach achieved an average accuracy of 96.8\%. The best results were obtained by integrating VGG-19 features with handcrafted features. Aljuaid et al. introduced a computer-aided diagnosis system for the BCC through deep neural networks (DNNs), transfer learning, and different magnification factors, coupled with data augmentation \cite{aljuaid2022computer}. In their research, there were three DNNs applied for classification. ResNet had the highest average accuracy, at 97.81\% for multi-classes.

\section{Methodology}
This section describes the methodology adopted for classifying breast cancer histopathology images into binary and multi-class categories. The approach utilizes the EfficientNet model with global average pooling, dropout, and dense layers with appropriate activation functions for classification. EfficientNet is a specialized architecture within the broader family of CNNs. It has been selected for this work due to its structured approach toward scaling architectural parameters, hence increasing its adaptability for complex patterns in medical images. CNNs have been overall effective in image classification tasks owing to their capability to learn spatial hierarchies through convolutional layers. The main operation in CNNs is the convolution, defined as:
\begin{equation}
y=f(x*W+b) 
\end{equation}
where \(x\) is the input, \(W\) represents the filter or kernel, \(b\) is a bias term, and \(*\) denotes the convolution operation. This process extracts spatial features from images, which are then passed through activation functions like ReLU to introduce non-linearity.

EfficientNet builds upon CNN principles with an optimized scaling approach, referred to as compound scaling, which allows the model to scale up efficiently in three dimensions: depth, width, and resolution. Its compound scaling formula scales depth, width, and resolution in a balanced way, given by:
\begin{equation}
\text{Depth} = \alpha^\phi, \quad \text{Width} = \beta^\phi, \quad \text{Resolution} = \gamma^\phi
\end{equation}
where \( \alpha \), \( \beta \), and \( \gamma \) are the scaling factors (typically optimized through neural architecture search), and \( \phi \) is a compound coefficient that controls the overall scaling. Its compound scaling aims to approximately double the computational cost for each unit increase in \( \phi \), resulting in a constraint:
\begin{equation}
\alpha \cdot \beta^2 \cdot \gamma^2 \approx 2
\end{equation}
This approach enables EfficientNet to pay more attention to detail while still maintaining computational efficiency. Its architectural design (as shown in Figure 1) is especially well-suited for large datasets with intricate details, like histopathological images, as it tries to balance the model’s capacity for generalization with computational efficiency. Additional enhancements in our methodology involve data augmentation and transfer learning in overcoming one of the most common problems with almost any medical dataset - class imbalance. This increases the model’s performance across all classes by ensuring the model generalizes appropriately for balanced performance.
\subsection{Proposed Framework for Handling Data Imbalance}
Neural network data imbalance normally leads to biased model performance, where extremely underrepresented classes are not predicted very accurately, mainly because of the inadequate number of training examples \cite{chen2023pcct}. Two methods were integrated into our methodology to overcome this issue. First, an intensive data augmentation technique was applied to images from underrepresented classes, specifically those with fewer images than the mean number of images across all classes. This allowed more balanced representation during training, hence reinforcing the generalization and fairness of the model in classification outcomes. In addition, we applied cost-sensitive learning by assigning higher misclassification costs to the minority classes. Figure 1 illustrates the proposed system workflow. 
\begin{figure}[H]
    \centering
    \includegraphics[width=0.8\textwidth]{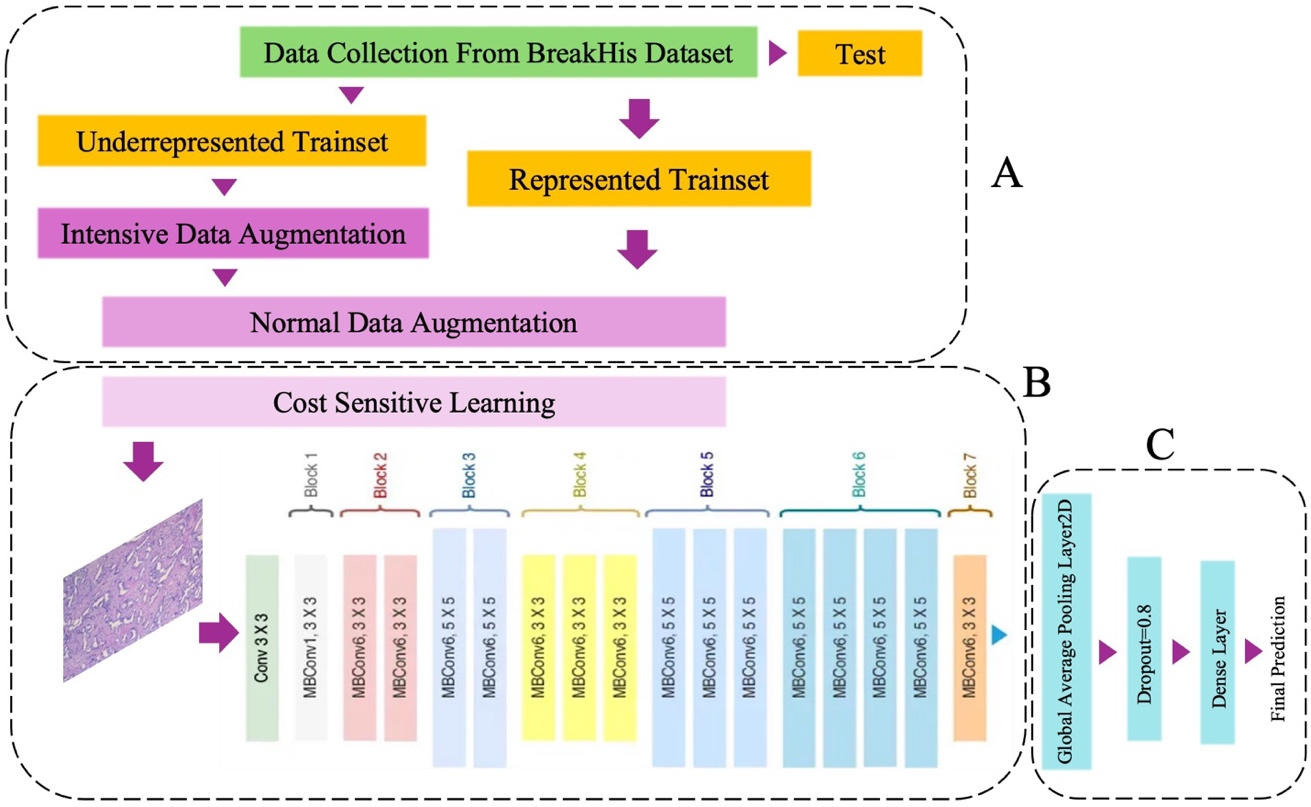} 
    \caption{Schematic of the proposed system workflow. Section A involves data preprocessing, including splitting the dataset into training and testing sets, followed by two levels of augmentation: intensive data augmentation for underrepresented classes and normal augmentation for the remaining data. Section B highlights the model architecture, utilizing EfficientNet with cost-sensitive learning. Section C covers the final layers of the model. }
    \label{fig:workflow}
\end{figure}
Combining these two approaches offers several advantages: it enhances data augmentation, and thus the model’s capability for generalization, and cost-sensitive learning adjusts the decision boundary of the model to be much more sensitive towards underrepresented classes \cite{domingos1999metacost}. This will reduce the possibility of overfitting in the majority class, hence providing more reliable and balanced predictions. These approaches put together provide class-balanced and accurate prediction models across all classes, hence increasing the general robustness and reliability of our model. The EfficientNet architecture is then utilized, followed by global average pooling, dropout, and dense layers as shown in Figure 1. The top layer is then customized to make use of a SoftMax activation function for multi-class classification and a sigmoid activation function for binary classification, returning the final predictions correspondingly.
\subsection{Data Augmentation}
Data augmentation is a common technique to enhance the diversity of training data, which helps reduce the risk of overfitting and improves model generalization. Standard augmentation methods used in breast cancer histopathology include rescaling, shearing, zooming, flipping, rotating, shifting, and adjusting brightness \cite{ijaz2023modality}, \cite{han2017breast}. We applied these augmentation strategies uniformly across all classes, using hyperparameter optimization to evaluate parameter values at three levels, as outlined in Table 1. According to the result of validation accuracy displayed in Figure 2, Level 2 provided the optimal balance, achieving the highest accuracy. This suggests that moderate augmentation yields the most effective improvement of generalization for our model.

\begin{table}[H]
\centering
\caption{Standard data augmentation parameters.}
\begin{tabular}{|c|c|c|c|}
\hline
Parameter & Level 1 & Level 2 & Level 3 \\ \hline
Rescale & 1./255 & 1./255 & 1./255 \\ \hline
Shear Range & 0.1 &	0.2 & 0.4  \\ \hline
Zoom Range & 0.1 & 0.2 & 0.4  \\ \hline
Horizontal Flip & True & True & True \\ \hline
Rotation Range & 20° & 30° & 40° \\ \hline
Width Shift Range & 0.1	& 0.2 & 0.4  \\ \hline
Height Shift Range & 0.1 & 0.2 & 0.4  \\ \hline
Brightness Range &	- &	[0.9, 1.1] & [0.9, 1.1] \\ \hline
\end{tabular}
\label{table:augmentation}
\end{table}
\begin{figure}[H]
    \centering
    \includegraphics[width=0.8\textwidth]{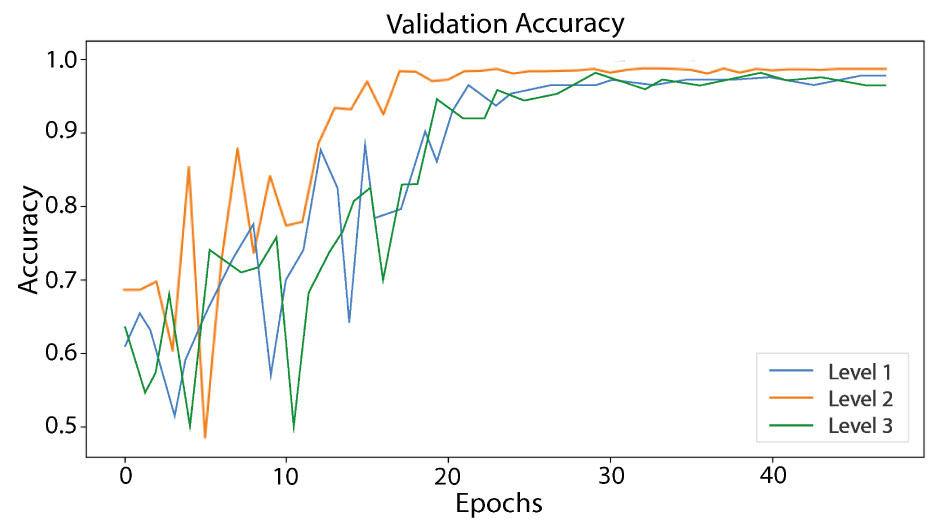} 
    \caption{Validation accuracy across three augmentation parameter intensities.}
    \label{fig:workflow}
\end{figure}
For underrepresented classes, a more intensive augmentation strategy is employed, as described in detail in Table 2. The proposed custom pipeline provides a very substantial increase in the size of the dataset in such classes by applying a set of aggressive transformations: flipping, affine transformation, adjustment of brightness, Gaussian blur, and additive Gaussian noise as shown in Figure 3. These operations were applied multiple times for every image to ensure better distribution of the dataset, which would result in a better generalization of the model to classify all classes correctly. Since the goal here is to enhance diversity in underrepresented classes rather than to fine-tune specific parameters, extensive hyperparameter optimization was not necessary.
\begin{table}[H]
\centering
\caption{Intensive data augmentation parameters}
\begin{tabular}{|c|c|}
\hline
Parameters & Values \\ \hline
Flip Horizontal & 50\% probability \\ \hline
Flip Vertical & 20\% probability \\ \hline
Affine Transformations & Rotation between -45 and 45 degrees \\ \hline
Brightness Adjustment & Multiply factor between 0.8 and 1.2 \\ \hline
Gaussian Blur & Sigma between 0.0 and 3.0 \\ \hline
Additive Gaussian Noise & Scale between 0.01255 and 0.05255 \\ \hline
\end{tabular}
\label{table:augmentation}
\end{table}
\begin{figure}[H]
    \centering
    \includegraphics[width=0.8\textwidth]{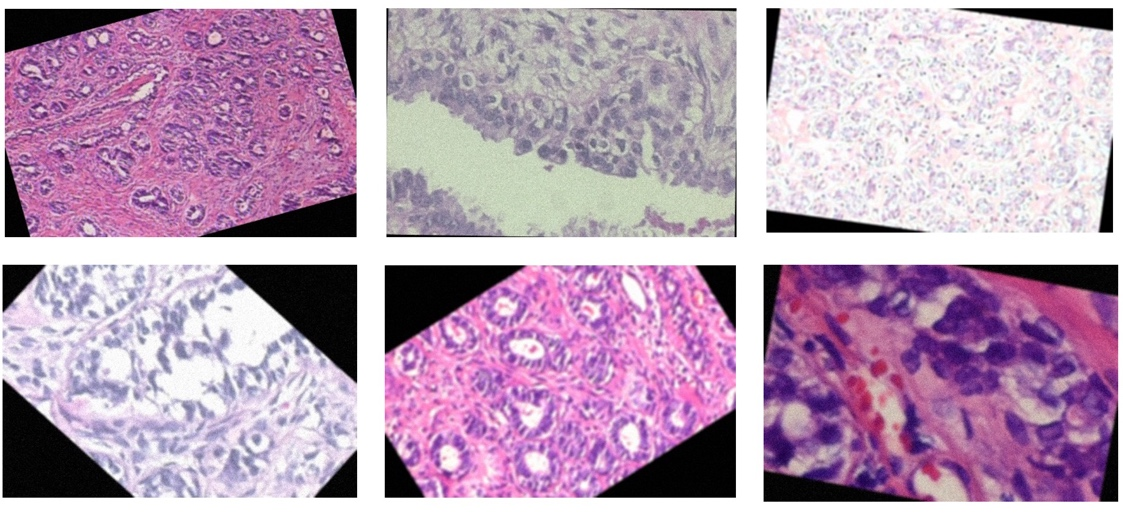} 
    \caption{Augmented histopathological breast tissue images. The first row displays Adenosis and Phyllodes Tumor images, and the second row shows Tubular Adenoma images, each subjected to intensive augmentation techniques, including horizontal and vertical flips, affine transformations, brightness adjustment, Gaussian blur, and additive Gaussian noise.}
    \label{fig:workflow}
\end{figure}
\subsection{Transfer Learning}
Transfer learning is one of the most potent techniques in image classification, improving the performance of deep learning networks on specific tasks. In this study, we applied EfficientNet B5 [13], and a transfer learning model for classifying breast cancer histopathology images. EfficientNet B5, known for its balanced scaling of depth, width, and resolution, serves as the backbone of our model. Initially, a binary classification model was trained to be used as a baseline for multi-class classification. Then, the learned weights of this binary model were fine-tuned for multi-class classification instead of using imageNet weights directly, allowing it to leverage the initially learned weights while effectively adapting to increased complexity in distinguishing between multiple classes. This approach is chosen for the following reasons:

\hangindent=1cm
•	The binary model pre-trained on ImageNet has already adapted 
the weights to some specific domains of histology images. That means the model has learned features relevant to the dataset, providing a better starting point for the multi-class task compared to more general features learned from ImageNet.

\hangindent=1cm
•	Fine-tuning model that has already been trained on a similar 
    task (binary classification on this dataset) can expedite training for the multi-class task. Since the model is already familiar with key features within this specific dataset, it requires less time and data to achieve high accuracy.

\hangindent=1cm
•	The binary classification can serve as a kind of intermediate 
    transfer learning, helping the model to better understand the nuances of our specific images. This appropriate understanding can result in better performance on the multi-class task, especially when the classes in the multi-class task are associated with the distinctions made in the binary task, that somehow have hierarchical or interrelated structure.
    
The code for our implementation is publicly available at \href{https://github.com/majid9418/Breast-Tumor-Classification-Histopathological}{GitHub}.

\section{Results and Discussions}
We presented our investigation’s results, evaluating our model’s efficacy on binary and multi-class classification problems. The discussion focuses on the efficiency gained from model architecture and proposed approaches for handling the class imbalance problem. The detailed discussions are made to outline the model’s capability in distinguishing between benign and malignant cases, as well as its generalization with our pre-trained weights to more complex multi-class classifications along with using pre-trained weights from our earlier binary classification model.

This study was conducted on Google Colab using Python 3.9 with TensorFlow 2.12 and Keras. The Colab environment provided access to NVIDIA Tesla T4 GPUs, offering sufficient computational power for training deep learning models. The model compilation utilized the 'adam' optimizer with a categorical cross-entropy loss function for multi-class classification, running under CUDA 11.2 to accelerate GPU processing. Additionally, Google Colab’s backend facilitated memory and processing optimizations necessary for handling large histopathological image datasets.
\subsection{Model Evaluation}
The findings are discussed in terms of accuracy, precision, recall, and F1-score with comparisons made to existing methods where applicable. Besides, there will be different sets of observations and predictions that are usually referred to as True Positive (TP), True Negative, (TN) False Positive (FP), and False Negative (FN). Using these metrics can determine all the measurements for computing the success rate, and those metrics are listed in Table 3.
\begin{table}[h]
\centering
\caption{Evaluation Metrics}
\begin{tabular}{|c|c|}
\hline
Metrics & Mathematical formulas \\ \hline
Accuracy & (TP + TN) / (TP+TN+FP+FN) \\ \hline
Precision & (TP / (TP+FP) \\ \hline
Recall & (TP / (TP+FN) \\ \hline
F1-Score & 2 × (Precision × Recall)/(Precision + Recall) \\ \hline
\end{tabular}
\label{table:Evaluation metric}
\end{table}

The original BreakHis dataset consists of 7,909 histopathological images of breast tissue from both benign and malignant tumor classes. We split the dataset into three parts: 80\% for training, 10\% for validation, and 10\% for testing. Initially, the distribution in the training set was 1,984 benign and 4,343 malignant images, indicating a significant class imbalance problem. In this regard, we adopted intensive augmentation only on the underrepresented benign class in the training set, increasing its size to 3,968 images, while the malignant class remained at 4,343. The augmentation methodology improved the distribution of the training set, enhancing the model to learn discriminative features from both classes. For multi-class classification, similar augmentation techniques were applied to address imbalance issues in six underrepresented subclasses, as detailed in Table 4.
\begin{table}[h!]
    \centering
    \caption{Training set distribution of BreakHis dataset after intensive data augmentation.}
    \begin{tabular}{|l|l|c|c|}
    \hline
    \textbf{Classes} & \textbf{Subclasses} & \textbf{Original Number} & \textbf{Final Number} \\ \hline
    Benign & Adenosis (A) & 355 & 710 \\ \hline
    Benign & Fibroadenoma (F) & 811 & 811 \\ \hline
    Benign & Tubular Adenoma (TA) & 455 & 724 \\ \hline
    Benign & Phyllodes Adenoma (PA) & 362 & 910 \\ \hline
    Malignant & Ductal Carcinoma (DC) & 2760 & 2760 \\ \hline
    Malignant & Lobular Carcinoma (LC) & 500 & 1000 \\ \hline
    Malignant & Mucinous Carcinoma (MC) & 633 & 1266 \\ \hline
    Malignant & Papillary Carcinoma (PC) & 448 & 896 \\ \hline
    \multicolumn{2}{|l|}{\textbf{Total}} & \textbf{6324} & \textbf{9077} \\ \hline
    \end{tabular}
    \label{tab:class_distribution}
\end{table}
In the binary classification task, as shown in Table 5, the proposed framework that addresses data imbalance outperformed the standard EfficientNet B5 model by yielding higher accuracy with a significant improvement in Precision and Recall. In the benign class, the Recall improved from 0.92 to 0.95 and the accuracy increased from 97.35\% to 98.23\%. These results, reflected in the confusion matrix Figure 4, indicate that the proposed framework handles data imbalance for more coherent and reliable classification compared to the standard augmentation techniques, especially in underrepresented classes.
\begin{table}[ht]
    \centering
    \caption{Binary classification performance comparison between EfficientNet B5 and the proposed framework (font in bold indicates better results).}
    \begin{tabular}{|l|l|c|c|c|c|c|}
    \hline
    \textbf{Model}            & \textbf{Class} & \textbf{Precision} & \textbf{Recall} & \textbf{F1-score} & \textbf{Support} & \textbf{Accuracy} \\ \hline
    EfficientNet B5           & Benign         & 0.99               & 0.92            & 0.96              & 248              & 97.35\%           \\ \cline{2-6}
                              & Malignant      & 0.97               & 1.00            & 0.98              & 543              &                   \\ \hline
    Proposed Framework        & Benign         & 0.99               & \textbf{0.95}   & \textbf{0.97}     & 248              & \textbf{98.23\%}  \\ \cline{2-6}
                              & Malignant      & \textbf{0.98}      & 1.00            & \textbf{0.99}     & 543              &                   \\ \hline
    \end{tabular}
    \label{tab:model_comparison}
\end{table}
\begin{figure}[H]
    \centering
    \includegraphics[width=0.8\textwidth]{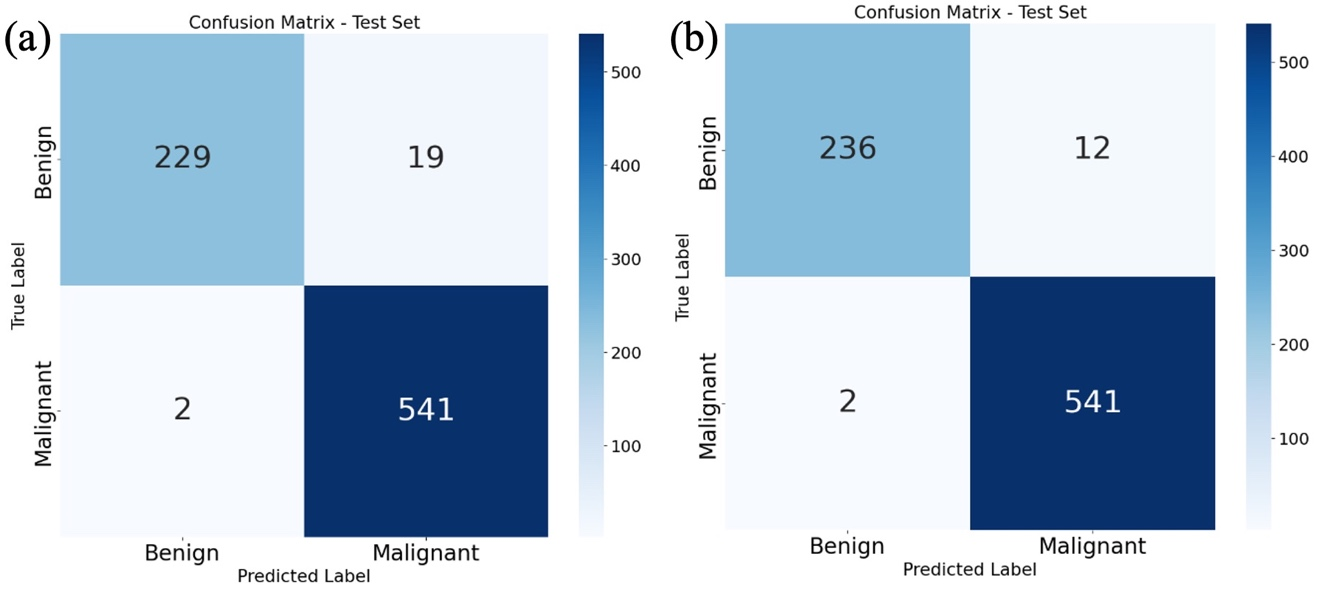} 
    \caption{Confusion matrices for binary classification results. EfficientNet B5 (a) without and (b) with intensive augmentation.}
    \label{fig:Binary matrix}
\end{figure}

The results in the case of the multi-class classification task were greatly improved as we further worked with advanced techniques shown in Table 6, Table 7, and Table 8. The overall accuracy was 91.27\% when only normal data augmentation was employed - see Table 6. With aggressive augmentation applied (see Table 7), the performance of the model significantly improved, reaching an accuracy of 94.54\%, improving further to 95.04\% when transfer learning was applied, as shown in Table 8 and its confusion matrix (Figure 7).

For individual classes, the intensive augmentation and transfer learning approach showed tremendous improvements. Precision for Adenosis improved from 0.98 (with intensive augmentation as shown in Table 7) to a perfect 1.00 across the board, as illustrated in the confusion matrix of Figure 5. For Ductal Carcinoma, initial misclassifications under normal augmentation—primarily confusing some cases with Lobular and Mucinous Carcinoma—were substantially reduced through intensive augmentation and transfer learning, as shown in the confusion matrix of Figure 6. This led to an increase in recall from 0.86 to 0.92. For Fibroadenoma, both precision and recall showed significant improvement with intensive augmentation, rising from 0.90 and 0.99 to 0.96 and 1.00, respectively (Tables 6 \& 7), and maintained these levels with transfer learning as shown in Table 8. Phyllodes Tumor performance also improved notably; recall increased from 0.84 under normal augmentation to 0.96 with intensive augmentation. Initially, seven instances were misclassified as Fibroadenoma (Figure 5), but with intensive augmentation, this dropped to only two misclassifications, as seen in Figure 6, with transfer learning sustaining this improved rate.

With intensive augmentation and transfer learning, the proposed method offers significant improvement in general for classes with a low number of instances, such as Adenosis, Fibroadenoma, and Phyllodes Tumor. This tends to indicate that such designed strategies are quite effective in improving results in the classification of multiple histopathological classes.
\begin{table}[ht]
\centering
\caption{Multi-class classification performance with normal augmentation, based on the original distribution of the dataset. Precision, recall, F1-score, and support are provided for each subclass.}
\begin{tabular}{|l|c|c|c|c|c|}
\hline
\textbf{Subclass} & \textbf{Precision} & \textbf{Recall} & \textbf{F1-score} & \textbf{Support} & \textbf{Accuracy} \\ \hline
Adenosis & 0.98 & 1.00 & 1.00 & 44 & \multirow{8}{*}{91.27\%} \\ \cline{1-5}
Ductal Carcinoma & 0.98 & 0.86 & 0.91 & 345 & \\ \cline{1-5}
Fibroadenoma & 0.90 & 0.99 & 0.94 & 101 & \\ \cline{1-5}
Lobular Carcinoma & 0.62 & 0.92 & 0.74 & 63 & \\ \cline{1-5}
Mucinous Carcinoma & 0.95 & 0.94 & 0.94 & 79 & \\ \cline{1-5}
Papillary Carcinoma & 0.86 & 0.98 & 0.92 & 56 & \\ \cline{1-5}
Phyllodes Tumor & 1.00 & 0.84 & 0.92 & 45 & \\ \cline{1-5}
Tubular Adenoma & 0.98 & 0.98 & 0.98 & 57 & \\ \hline
\end{tabular}
\label{tab:subclass_metrics}
\end{table}
\begin{table}[ht]
\centering
\caption{Multi-class classification performance after adopting intensive data augmentation.}
\begin{tabular}{|l|c|c|c|c|c|}
\hline
\textbf{Subclass} & \textbf{Precision} & \textbf{Recall} & \textbf{F1-score} & \textbf{Support} & \textbf{Accuracy} \\ \hline
Adenosis & \textbf{1.00} & 1.00 & 1.00 & 44 & \multirow{8}{*}{\textbf{94.54\%}} \\ \cline{1-5}
Ductal Carcinoma & 0.96 & \textbf{0.92} & \textbf{0.94} & 345 & \\ \cline{1-5}
Fibroadenoma & \textbf{0.96} & \textbf{1.00} & \textbf{0.98} & 101 & \\ \cline{1-5}
Lobular Carcinoma & \textbf{0.66} & 0.79 & 0.72 & 63 & \\ \cline{1-5}
Mucinous Carcinoma & \textbf{1.00} & \textbf{0.97} & \textbf{0.99} & 79 & \\ \cline{1-5}
Papillary Carcinoma & \textbf{0.98} & \textbf{1.00} & \textbf{0.99} & 56 & \\ \cline{1-5}
Phyllodes Tumor & 1.00 & \textbf{0.96} & \textbf{0.98} & 45 & \\ \cline{1-5}
Tubular Adenoma & \textbf{1.00} & 0.98 & \textbf{0.99} & 57 & \\ \hline
\end{tabular}
\label{tab:subclass_metrics_updated}
\end{table}
\begin{table}[ht]
\centering
\caption{Performance metrics for multi-class classification utilizing intensive data augmentation combined with our transfer learning approach.}
\begin{tabular}{|l|c|c|c|c|c|}
\hline
\textbf{Subclass} & \textbf{Precision} & \textbf{Recall} & \textbf{F1-score} & \textbf{Support} & \textbf{Accuracy} \\ \hline
Adenosis & 1.00 & 1.00 & 1.00 & 44 & \multirow{8}{*}{\textbf{95.04\%}} \\ \cline{1-5}
Ductal Carcinoma & \textbf{0.97} & 0.92 & 0.94 & 345 & \\ \cline{1-5}
Fibroadenoma & 0.96 & 1.00 & 0.98 & 101 & \\ \cline{1-5}
Lobular Carcinoma & \textbf{0.67} & \textbf{0.80} & \textbf{0.73} & 63 & \\ \cline{1-5}
Mucinous Carcinoma & 1.00 & 0.97 & 0.99 & 79 & \\ \cline{1-5}
Papillary Carcinoma & 0.98 & 1.00 & 0.99 & 56 & \\ \cline{1-5}
Phyllodes Tumor & 1.00 & 0.96 & 0.98 & 45 & \\ \cline{1-5}
Tubular Adenoma & 1.00 & 0.98 & 0.99 & 57 & \\ \hline
\end{tabular}
\label{tab:subclass_metrics_new}
\end{table}
\begin{figure}[H]
    \centering
    \includegraphics[width=0.5\textwidth]{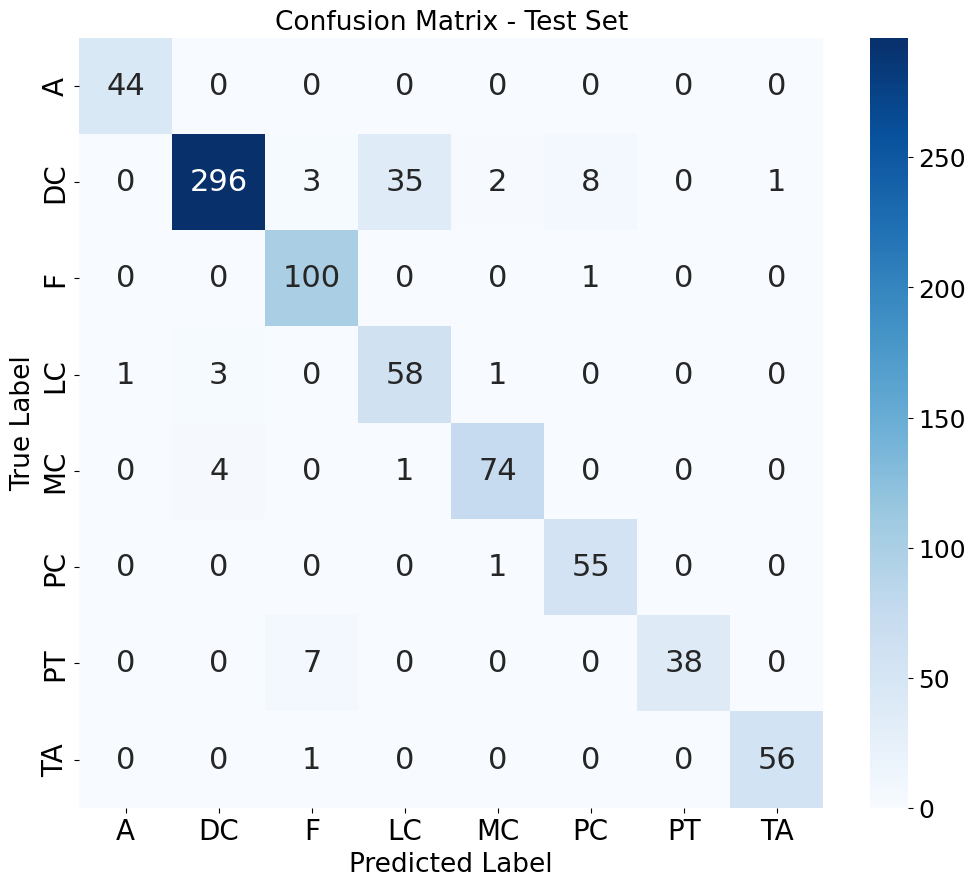} 
    \caption{Confusion matrix illustrating the results of multi-class classification with normal data augmentation.}
    \label{fig:multi_matrix_1}
\end{figure}

\vspace{-10pt} 

\begin{figure}[H]
    \centering
    \includegraphics[width=0.5\textwidth]{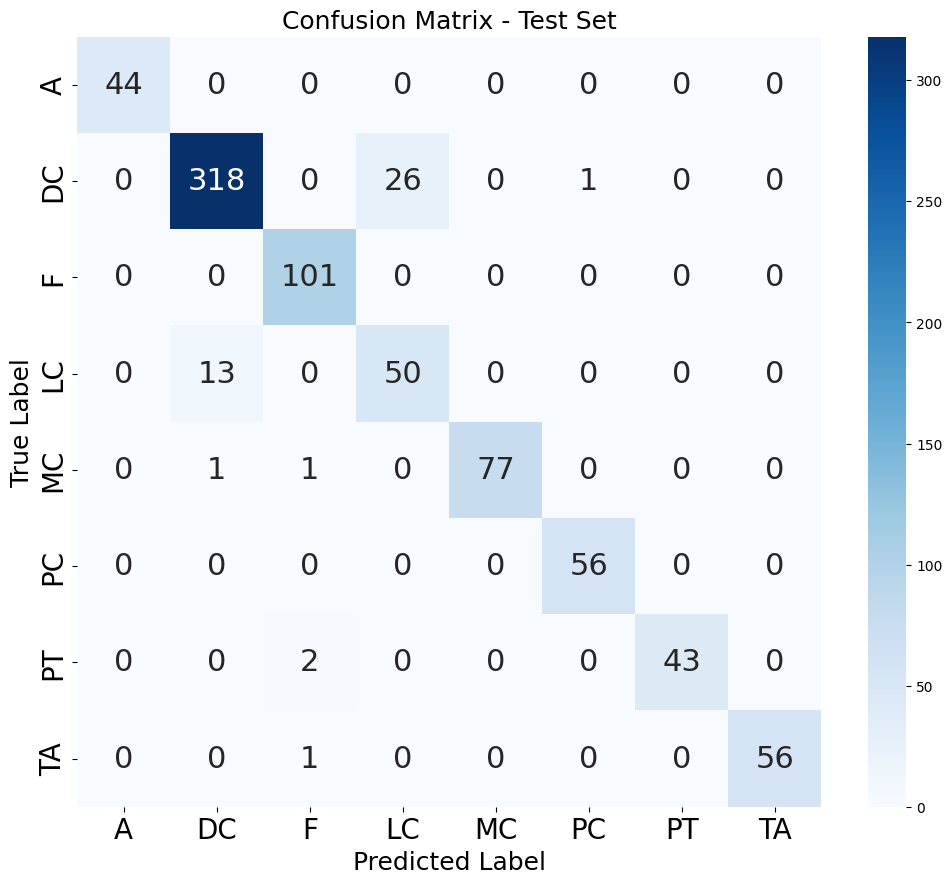} 
    \caption{Confusion matrix illustrating the results of multi-class classification with intensive data augmentation.}
    \label{fig:multi_matrix_2}
\end{figure}

\vspace{-10pt} 

\begin{figure}[H]
    \centering
    \includegraphics[width=0.5\textwidth]{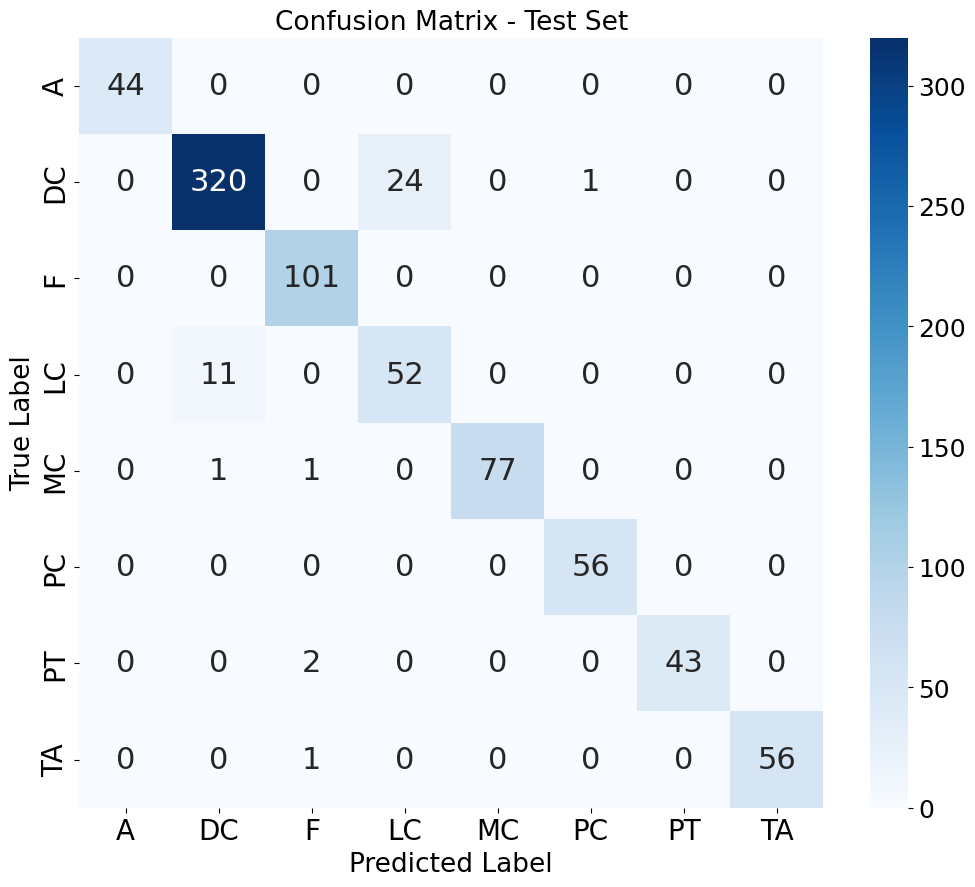} 
    \caption{Confusion matrix illustrating the results of multi-class classification with intensive data augmentation and our transfer learning method.}
    \label{fig:multi_matrix_3}
\end{figure}
\subsection{Comparison With Other Works}
As summarized in Table 9, which compares the proposed framework with other state-of-the-art methods for breast tumor classification, our EfficientNet-based workflow outperforms existing approaches, particularly in multi-class classification tasks. For instance, while models such as MSIMFNet and CSDCNN achieve test accuracies of 88.00\% and 93.32\%, respectively, our framework elevates accuracy to 95.04\% with transfer learning. This improvement is likely due to our model’s ability to better handle class imbalance and apply a more advanced augmentation strategy, resulting in more consistent and reliable predictions across underrepresented classes.
Additionally, our framework demonstrated strong performance in binary classification, achieving a test accuracy of 98.23\%, further underscoring its robustness across various classification tasks. This highlights the potential of our workflow for generalization beyond the EfficientNet architecture, as it can be adapted to enhance classification performance in other models, particularly in cases of data imbalance and complex class structures.
\begin{table}[ht]
\centering
\caption{Comparative performance analysis of breast tumor classification methods.}
\begin{tabular}{|l|c|c|c|c|}
\hline
\textbf{Method} & \multicolumn{2}{c|}{\textbf{Test Accuracy}} & \multicolumn{2}{c|}{\textbf{Validation Accuracy}} \\ \cline{2-5} 
                & \textbf{Binary} & \textbf{Multi-class} & \textbf{Binary} & \textbf{Multi-class} \\ \hline
DenseNet \cite{man2020classification}        & 92.21\%       & 87.30\%        & -              & -             \\ \hline
MSIMFNet \cite{sheikh2020histopathological}        & 97.32\%       & 88.00\%        & -              & -             \\ \hline
SECS \cite{yu2023secs}            & -             & -              & 99.75\%        & 95.69\%       \\ \hline
DRDA-Net \cite{chattopadhyay2022drda}        & 96.97\%       & -              & -              & -             \\ \hline
CSDCNN \cite{han2017breast}          & 96.07\%       & 93.32\%        & -              & -             \\ \hline
VGG-19 and Handcrafted [10] & -             & -              & 99.70\%        & 96.80\%       \\ \hline
GLNet \cite{khan2024glnet}           & -             & 91.21\%        & -              & -             \\ \hline
EfficientNet (This Work) & \textbf{98.23\%}       & \textbf{95.04\%}        & 99.12\%        & 99.25\%       \\ \hline
\end{tabular}
\label{tab:method_comparison}
\end{table}
\section{Conclusion}
This work presents a novel framework for improving the classification performance of breast cancer histopathological images, especially addressing important challenges such as class imbalance and model fine-tuning. EfficientNet-B5 was first trained using standard augmentation and cost-sensitive learning to establish a baseline performance. Class imbalance was then mitigated through intensive data augmentation, which yielded significant improvements in the classification accuracy of minority classes. In addition, we fine-tuned our multi-class classification model by utilizing transfer learning on the pre-trained weights of the binary classification task. This results in an overall massive gain in classifying the various subclasses of breast cancers. 
    The key contributions of this study are as follows: (1) EfficientNet has been successfully applied, for the first time, to classify histopathological images of breast cancer, demonstrating robustness and effectiveness in this domain. (2) A comprehensive evaluation of data augmentation and class balancing strategies was conducted, proving these methods’ efficacy in improving model performance. (3) Transfer learning was implemented by fine-tuning the binary classification model, EfficientNetB5, for multi-class classification, leveraging learned features to achieve high accuracy in distinguishing breast cancer histology subclasses. (4) Comparative analysis with state-of-the-art models confirms that the proposed approach surpasses existing methods, especially in handling class imbalance and improving classification accuracy.
    
    This methodology not only outperforms current techniques but also offers a flexible framework adaptable to other models or datasets with similar classification challenges. Future work may include using GANs to generate high-fidelity images for underrepresented categories, further enhancing model robustness.
\section*{Acknowledgments}
K. W. acknowledges the financial support by Texas Tech University through HEF New Faculty Startup, NRUF Start Up, and Core Research Support Fund.

\clearpage
\bibliography{references.bib}
\bibliographystyle{plain}

\end{document}